\documentclass[journal=jacsat,manuscript=article]{achemso}

\usepackage[version=3]{mhchem}
\usepackage{amsmath}
\usepackage{graphicx}
\usepackage{svg}
\usepackage{soul}
\usepackage{xr}

\author{Xiaohe Lei}
\affiliation{Department of Chemistry and Biochemistry, University of California, Santa Barbara, Santa Barbara, California 93106, United States}
\author{Annabelle Canestraight}
\affiliation{Department of Chemical Engineering, University of California, Santa Barbara, Santa Barbara, California 93106, United States}
\author{Vojtech Vlcek}
\email{vlcek@ucsb.edu}
\affiliation{Department of Chemistry and Biochemistry, University of California, Santa Barbara, Santa Barbara, California 93106, United States}
\alsoaffiliation{Department of Materials, University of California, Santa Barbara, Santa Barbara, California 93106, United States}
\title[]
 {Exceptional spatial variation of charge injection energies on plasmonic surfaces} 

\begin{document}

\clearpage
\begin{abstract}
Charge injection into a molecule on a metallic interface is a key step in many photo-activated reactions. The energy barrier for injection is paralleled with the lowest particle and hole addition energies. We employ Green's function formalism of the many-body perturbation theory and compute the excitation energies, which include non-local correlations due to charge density fluctuations on the surface, i.e., the plasmons. We explore a prototypical system: \ce{CO_2} molecule on nanoscale plasmonic Au infinite and nanoparticle surface with nearly 3,000 electrons. In contrast to widely used density functional theory, we demonstrate that the energy barrier varies significantly depending on the molecular position on the surface, creating ``hot spots'' for possible carrier injection. These areas arise due to an intertwined competition between purely plasmonic couplings (charge density fluctuations on the substrate surface alone) and the degree of hybridization between the molecule and the substrate. There are multiple positions found with the lowest energy barrier for the electron/hole injection. We identify that the charge injection barrier to the adsorbate on the plasmonic surface trends down from the facet edge to the facet center -- here, the change in molecular orbitals overshadows the role of the charge fluctuations in the substrate. This finding contrasts the typical picture in which the electric field enhancement on the nanoparticle edges is considered the most critical factor. 
\end{abstract}

Plasmon-driven interactions on metallic nanostructures are well-known for improving the sensitivity of the optical probes\cite{moskovits1978surface} and enabling light-driven reactions\cite{buntin1988optically,zhu2002electron,christopher2011visible,christopher2012singular,mukherjee2013hot,avanesian2014adsorbate,christopher2017hot,duchene2018hot,kim2018surface,seemala2019plasmon,abouelela2021review,martirez2021first,jiang2022plasmonic,lou2022direct,wang2023plasmonic}. The incident electromagnetic radiation induces localized surface plasmon resonance (LSPR) on the subwavelength metal surface, which leads to increased light absorption and scattering.\cite{wang2023plasmonic,martirez2021first,hernandez2018first, christopher2017hot}
The LSPR coupling to the adsorbate can trigger chemical reactions that are hard to achieve by thermal excitation (e.g., water splitting\cite{abouelela2021review}) and lead to selective reaction pathways ( e.g., selective products formation in \ce{CO_2} reduction\cite{duchene2018hot,kim2018surface,jiang2022plasmonic,wang2023plasmonic}).
The interfacial energy transfer following the LSPR decay is critical to understanding such plasmon-mediated mechanism of chemical transformations.\cite{zhu2002electron} The LSPR dephasing is driven by multiple possible scattering channels, including a nonradiative pathway generating ``hot'' charge carriers, i.e., highly excited and out-of-equilibrium holes and electrons.\cite{buntin1988optically,wu2015efficient,avanesian2014adsorbate,kale2015plasmons} 
The efficiency of the interfacial energy transfer via charge carrier injection is affected by multiple factors such as the incident light wavelength, the composition, size, and shape of the nanostructure, the local electric field, and the availability of the adsorbate state. \cite{zhu2002electron,kelly2003optical,seemala2019plasmon,mukherjee2013hot,lou2022direct}.

The local electric field enhancement by LSPR is known to be an important factor for the design of plasmonic photocatalysts for small molecule activation. The LSPR strength varies spatially across the plasmonic surface. The field at the plasmonic nanostructure junction, or the edge, is typically more enhanced compared to other areas and hence it is determined to be the chemical ``hot spot", where the plasmonically driven transformations may preferentially occur\cite{christopher2012singular,seemala2019plasmon,jiang2022plasmonic}. However, the effective energy barrier for charge injection derives from the energy level of the adsorbate states. While critical, this aspect of plasmonic transformation hot spots has not been sufficiently studied. In practice, it requires a detailed microscopic description of the coupling between the excited state of the plasmonic substrate and the adsorbate at any given location of the surface. A computational investigation is well-positioned to address this question, where the orbital energies are found by considering the complex many-body interactions between the subsystems. Density Functional Theory (DFT) is used most frequently, analyzing the charge density distribution\cite{chernyshova2018origin} and energetics\cite{wu2021identification,egger2015reliable,biller2011electronic} in interfacial systems. These calculations are affordable for systems of thousands of electrons\cite{Millionelectrons}. However, DFT cannot provide excited state properties.\cite{martin2016interacting}

We overcome the limitations of DFT by employing many-body perturbation theory (MBPT) in the $GW$ approximation (GWA) to solve for the  energies of individual single excited electrons and holes ``dressed'' by their interactions with the environment (i.e., quasiparticles -- QPs)\cite{hedin1965new,martin2016interacting}. In GWA, the QP propagator, represented by the Green's function ($G$), is renormalized by non-local correlation effects stemming from charge density fluctuations on the molecule and the substrate. The screened Coulomb interaction ($W$) contains the non-local interactions with the underlying plasmonic ``reservoir'' which affects molecular QP energies.\cite{hedin1965new,vlvcek2018quasiparticle,martin2016interacting}  In this work we use the $G_0W_0$ approximation, where both the Green's function and the screened coulomb interaction are constructed using the DFT eigenstates. Rather than self-consistently solving for the Green's function of the true QP states, a one shot correction to the KS Hamiltonian is applied. This method has been shown to improve the predictions of both experimental band-gaps and molecular ionization potentials, making it a strong theoretical tool for treating a molecule-metal interface.\cite{VojtechStochGW,YangHedinShift,Kaplan2015,rignanese2001quasiparticle,tamblyn2011electronic} However, the scaling of this algorithm has prevented its use in studying realistic systems.\cite{pham2013g,neuhauser2014breaking} To remedy this, we use the stochastic $GW$ formalism which reduces the computational scaling and allows us to handle systems with thousands of electrons, enabling this study.\cite{VojtechStochGW,vlvcek2018swift} 
We investigate \ce{CO_2} on gold, a prototypical surface-adsorbate system where the small molecule acts as a probe of local coupling of an excited molecular state to the metal substrate. Large slabs and a full nanoparticle (NP) with thousands electrons are used as the gold substrate. We complement the light-enhancement-driven reactivity picture by identifying the ``hot spots" with the minimal energy barrier for (vertical) hot charge injection to the adsorbate. 
We study the molecular QP energies corresponding to the lowest hole for electron injection energy onto the adsorbate across varied positions on Au surfaces. The electron injection process corresponds to the vertical excitation in the proposed mechanism of the fist step in \ce{CO_2} reduction, forming the transient negative ion intermediate\cite{hori2008electrochemical,chernyshova2018origin,wang2023plasmonic}. 

Our results demonstrate a remarkable variation of the molecular QP energies depending on the exact position on the surface (here corresponding to various facets of an NP). Accounting for non-local many-body interactions leads to changes on the order of up to several hundreds of meV and significantly affects the predicted energy barrier for charge injection, i.e., the distance from the substrate Fermi level. Our results indicate that the plasmonically-driven chemical ``hot spots" are affected by a non-trivial interplay between the adsorbate's sensitivity to the underlying plasmonic field and the adsorbate orbital hybridization with the delocalized surface states. In contrast to the previous models which study solely the local field enhancement, \cite{seemala2019plasmon,jiang2022plasmonic} we find that the charge injection barrier \textit{increases} as the molecule approaches the nanostructure edges, which are known to exhibit strong plasmon dephasing and a relative abundance of free charges. 

We study the molecule in various positions on Au (111) and (100) surfaces of an infinite slab (generated by 2D periodic boundary conditions) and an Au$_{260}$ NP. The physisorbed \ce{CO_2} geometry is optimized using DFT applied to the molecule on a 3-layer Au surface using QuantumEspresso \cite{giannozzi2009quantum}. For selected geometries, we repeated the relaxation with both a thicker gold slab and a wider gold slab to verify that the molecular geometry is insensitive to the increased slab thickness or decreased molecular coverage on the surface (with $< 0.1\%$ the bond-lengths and angles variation). The geometry optimization is performed on all types of adsorption sites. For(111): fcc hollow site (f), hcp hollow site (h), atomic site (a) and bridge site (b); for (100): hollow site (h), atomic site (a), and bridge site (b). The simulation system name is labeled as the facet index followed by the site initial, e.g., $(111)_f$ for the \ce{CO_2} adsorbed at fcc hollow site on Au (111) surface. 
The optimized surface separation is $3.44$\AA~ on (111) and $3.24$\AA~ on (100), with variations across different sites being $<5\%$. The variation of the total energy is minimal ($<0.0005\%$  between the minimum of (111)$_f$ and its counterpart (100)$_h$). We use the optimized molecular geometry and molecule-surface separation in relaxed $(111)_f$ and $(100)_h$ for the corresponding lattice surface to make a consistent comparison among the computational results. In addition, we have investigated a relaxed structure of \ce{CO_2} on Au (111)/(100) interface ridge. Further details about the structural optimization are in the Supplementary Information (SI). 

While the molecular geometry is determined primarily by the near-surface Au atoms, the electronic structure of the composite system converges more slowly with the number of gold layers. The fully converged single-particle energies are obtained for slabs with 9 and 13 layers for the (111) and (100) surfaces. These are comparable in dimension to a truncated octahedron Au$_{260}$ NP with (111) and (100) facets (Fig.S3), which is a thermodynamically stable form of a gold NP.\cite{hernandez2018first} The molecular geometry and separation are taken from the optimized structure outlined above.

To define the electronic states associated with the adsorbate, we first construct a set of localized molecular states $\{\varphi\}$ constructed as a linear combination of unitary-transformed maximally localized states obtained with the Pipek-Mezey Wannier approach\cite{weng2022reduced}. The $\{\varphi\}$ states span the molecular subspace and conceptually correspond to the states in an isolated molecule in the gas phase\cite{GwenSpatialDecay,weng2021efficient}. We use this step to identify the canonical mean-field states, $\{\phi\}$ (i.e., the delocalized state computed by DFT for the entire molecule-surface system) that have primarily molecular character. We rank the states $\{\phi\}$ according to their projection values $|p_j|^2=|\langle\varphi_j|\phi\rangle|^2$ where $\varphi_j$ is a particular molecular state.   
The $|p|^2$ values for the frontier molecular orbitals (FMOs) are $< 0.95$ due to hybridization, while the bottom valence states exhibit $|p|^2$ very close to unity. Upon adsorption, the interaction with the substrate lifts the degeneracy of the \ce{CO_2} highest occupied state (HOMO). The canonical states with the largest fraction of localized HOMO orbital is considered to be representative of the HOMO state on the surface. For the LUMO, we inspect only the bound states. These have  $|p|^2$ of $\approx 0.3$ at most and thus show more pronounced hybridization with the substrate.

We calculate the QP energies of injected hole and electron for these most molecule-like canonical orbitals. Employing the $G_0W_0$ approximation, the QP energy is found through the fixed point equation,
\begin{equation}
\epsilon_{QP} = \epsilon_{KS}+\langle\phi|\hat{\Sigma}_P(\omega=\epsilon_{QP})+\hat{\Sigma}_X-\hat{v}_{xc}|\phi\rangle
\end{equation}
where $\epsilon_{KS}$ is the DFT starting point eigenvalue, $\hat{v}_{xc}$ is the exchange-correlation potential (approximated by a semilocal DFT functional) and $\hat{\Sigma}_X$ is the non-local exchange self-energy and $\hat{\Sigma}_P$ is the dynamical self-energy that accounts for the induced density-density interactions, i.e., it includes the plasmonic response of the substrate. 

We first investigate the charge injection to \ce{CO_2} adsorbed on infinite Au (111) and (100) surfaces. The QP energy of FMOs varies only by $0.4$ eV between the two surfaces. The HOMO QP energy is $-12.6$ eV on average; this agrees well with the experimental result estimating the center of HOMO state near $-12.5$ eV.\cite{atkinson1974ultra} The LUMO is unbound in the gas phase, its average value on an infinite surface is $-1.10$ eV. The LUMO ``stabilization'' (negative energy shift making the orbital bound) due to the surface interaction thus, in principle, enables the CO$_2^-$ intermediate formation important in CO$_2$ reduction.
MBPT is known to correct the DFT results by large shifts\cite{onida2002electronic,blase2011first,rostgaard2010fully,rignanese2001quasiparticle,tamblyn2011electronic}, this is also the case here: For the HOMO, the $G_0W_0$ correction reduces the energy by $3.34$ eV and $3.64$ eV in $(111)_f^{inf}$ and $(100)_h^{inf}$.
For the LUMO, the QP energy is, on average, $0.38$ eV lower than the KS eigenvalue. As a result, we can clearly see that MBPT leads to a significant fundamental band gap opening compared to the DFT results by $35\%$ in $(111)_f^{inf}$ and $40\%$ in $(100)_h^{inf}$ (i.e., $2.86$ eV and $3.35$ eV).

We now investigate the charge injection onto the molecule on the Au$_{260}$ NP. The NP exhibits metallic behavior and supports surface plasmon resonance\cite{hernandez2018first}. Since it contains a similar number of Au atoms as the 2D infinite slab, the observed change in the non-local correlations should be influenced primarily by the NP shape. There are 8 layers for the Au (111) facet and 9 layers for the Au (100) facet in the NP. Yet, the molecular QP energies on the NP are significantly different from those on an infinite substrate, demonstrating the profound effects of nanoscale surface structure. On average, the QP energies on the NP are higher by $3.11$ eV for the hole and $0.70$ eV for the electron injection than that of the infinite surface result. The electron injection barrier from the substrate to the \ce{CO_2} $E_{QP}^{FMO}-E_f$ is shown to be higher than the simple difference $E_{KS}^{FMO}-E_f$ predicted by DFT. DFT underestimates the energy barrier for electron injection by $20\%$ and for hole injection by $10\%$. 

\begin{figure}[htbp]
        \centering
        \includegraphics[width=0.8\textwidth]{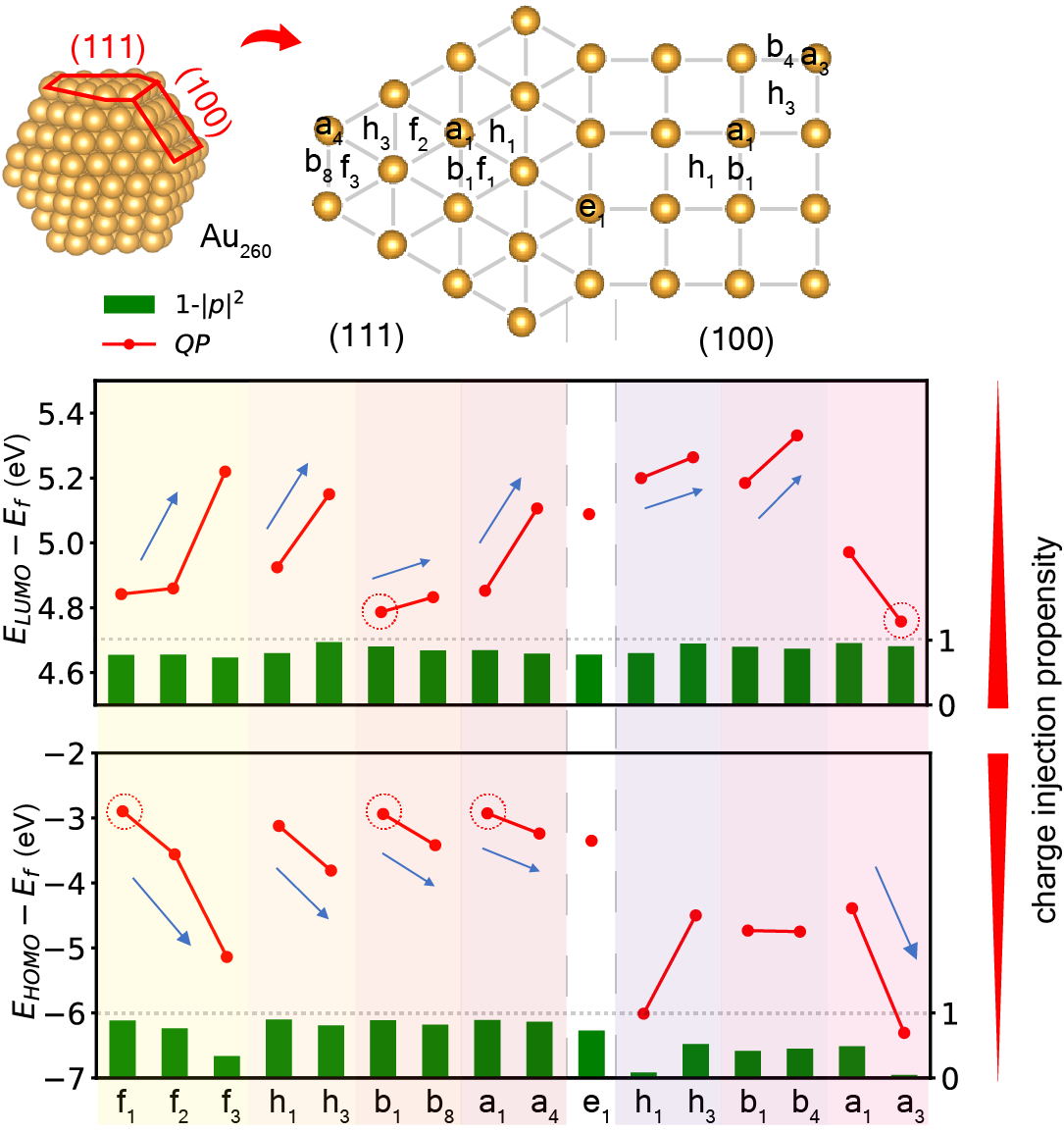}
        \caption{Au$_{260}$ NP consists of 8 hexagon (111) facets and 6 square (100) facets. The surface site map of the NP is displayed on its right. Four types of adsorption sites on (111) facet are: fcc hollow (f), hcp hollow (h), atomic (a), and bridge (b); Three types of adsorption sites on (100) are: hollow (h), atomic (a), and bridge (b). The positions are numbered from small to large, corresponding from the facet center position to the edge position. Labels for the symmetrical positions are not repeated. The case \ce{CO_2} is put atop the ridge Au atom is labeled as e1. In the QP energy panels, the difference between the QP energy and the Fermi level (red dots) is plotted together with the degree of hybridization $1-|p|^2$ values (green bars) for the selected hybridized FMO. Blue arrows indicate the QP energy trend further from the Fermi level when the molecular position approaches the NP facet edge. The data points corresponding to the lowest charge injection barriers are red-circled. the most favored sites for the electron injection are $(111)_{b1}$ and $(100)_{a3}$, with charge injection barrier at $4.76 \pm 0.07$ eV; the most favored sites for the hole injection are $(111)_{f1}$, $(111)_{b1}$ and $(111)_{a1}$, with the charge injection barrier at $2.90 \pm 0.08$ eV.} 
        \label{fig:QPvsKS}
    \end{figure}

These average results, however, smear out the local variation of the barriers. The most surprising discovery is that the QP energy varies significantly depending on the specific position of the molecule on the surface. As shown in Fig.\ref{fig:QPvsKS} (Top panel), for each type of surface site (namely on (111) facet: f, h, b, a; on (100) facet: h, b, a), we studied at least one facet center position and one facet edge position.
The difference between the center and the edge positions for the LUMO QP energy are as large as $380$ meV. Such difference corresponds to $1.5kT$ energetic difference at 298K. Considering the non-equilibrium process involving highly excited ``hot'' electron injection (and corresponding Boltzmann factors), this implies the injected electron on the NP facet center is about four folds ``hotter'' than it on the NP facet edge. The variation is even more pronounced for HOMO (i.e., and the corresponding hole injection barrier) that varies as much as $3.41$ eV (Fig.\ref{fig:QPvsKS}). These MBPT results are in striking contrast to mean-field calculations, which show $700\%$ smaller position dependence. The discrepancy is due to the substrate-induced renormalization, which is missing from the approximate (semilocal) exchange-correlation functional.\cite{kummel2008orbital,neaton2006renormalization,egger2015reliable,tamblyn2011electronic} 
 Our analysis shows that the hole injection is most probable in the center of $(111)$ facet, with the QP energy level of $2.90 \pm 0.08$ eV below the Fermi level. The electron injection is also most probable at the center of (111), but we also identified $(100)_{a3}$ edge position to host similarly low QP LUMO energy, with a QP energy of $4.76 \pm 0.07$ eV above the Fermi level. The overall injection energy variation is clearly related to the surface position.

\begin{figure}[htbp]
        \centering
        \includegraphics[width=\textwidth]{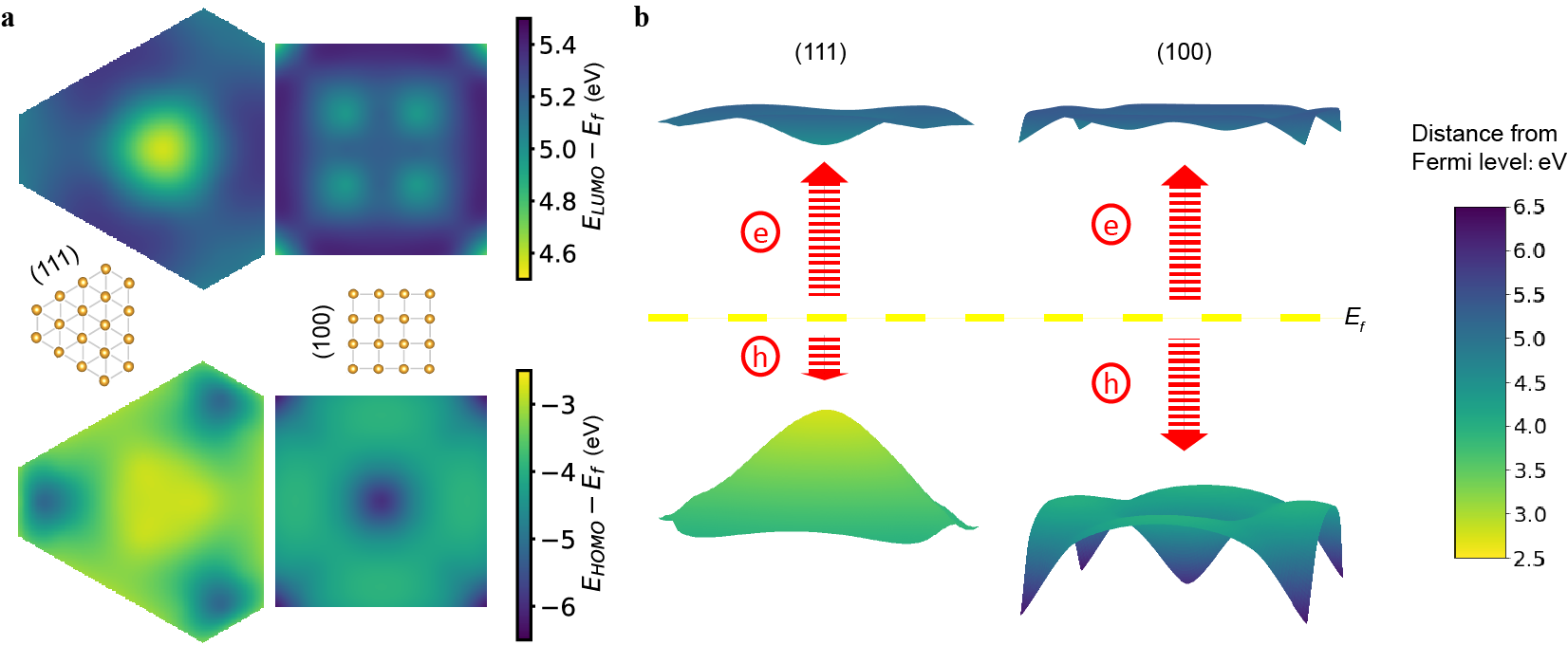}
        \caption{\textbf{(a)} Interpolated heat map of $E_{FMO}-E_{f}$ on (111) and (100) facets. The yellow end indicates a smaller injection barrier. \textbf{(b)} Interpolated surface plot of $E_{FMO}-E_{f}$ on (111) and (100) facets intuitively displaying exceptional spatial variation of the charge injection energies. A yellow dash line shows the Fermi level $E_f = 5.42$ eV. On the (111) facet, the center region is indicated to be the most favored location for both quasi-electron and quasi-hole injections. On (100) facet, the facet corner is indicated to be the most favored site for electron injection; the facet side location is indicated to be the most favored location for hole injection.} 
        \label{colormap}
    \end{figure}

As the interpolated color map shows (Fig.\ref{colormap}), on the NP (111) facet, both the electron and hole injection propensity decreases from the center to the edge. On the NP (100) facet, the center is no longer shown to be the most probable region for charge injection. The corner position is most favorable for the electron injection. As for the hole injection, the side location has the highest charge injection propensity.      
The spatial patterns of the charge injection barrier shown in Fig.\ref{colormap}(a) depend on multiple factors; here, the degree of hybridization with the surface should play a significant role as it affects the interaction with the surface states. We will now study the effect of orbital hybridization in greater detail by considering the molecular and substrate self-energy contributions separately. The QP energy shift evaluated for a given canonical state is decomposed into individual contributions based on the projection on the molecular and surface subspaces. For the analysis, we again employ a set of localized states and the values of $|p_j|^2$ defined above. The ``molecular contribution" to a particular state: is  $|p|^2 = \sum_j |p_j|^2$, while the remainder ($1-|p|^2$) is the surface contribution. For HOMO, the summation goes over all localized molecule-like occupied states $\{\varphi_j\}; j\le{\rm HOMO}$. We below also discuss the self-energy dependence on $|p_{\rm HOMO}|^2$ that exhibits the same behavior - illustrating that the exact definition of the molecular subspace is less critical. For the LUMO state, we restrict the projection only on a single lowest unoccupied state ($j=$LUMO), as we could not consistently identify another molecular state that would accept an extra electron. This simplistic approach shows that hybridization is a major factor: the spatial pattern of $1-|p|^2$ values (Fig.S5) is remarkably close to the maps shown in Fig.\ref{colormap}.

\begin{figure}[htbp]
        \centering
        \includegraphics[width=0.6\textwidth]{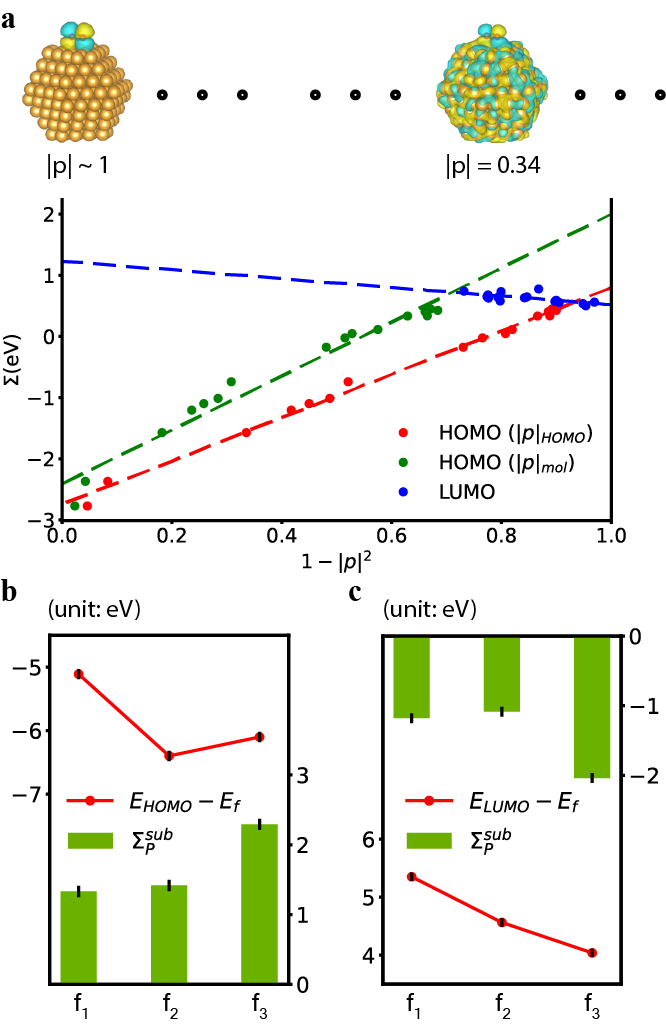}
        \caption{\textbf{(a)} The self-energy plot against the degree of hybridization $1-|p|^2$. Dash lines show the fitting curves in the simplified relation: $\Sigma = |p|^2 \Sigma_{\rm mol} + \left(1 - |p|^2\right) \Sigma_{\rm sub}$. Noting that the slopes $k=\Sigma_{sub}-\Sigma_{mol}$ shown by the HOMO fitting on $|p|^2=|p_{mol}|^2$ and $|p|^2=|p_{HOMO}|^2$ change much, but they exhibit the same dependence on the hybridization with the surface. On the top, the hybridization suppressed HOMO and the hybridized HOMO in (111)$_{f1}$ are exemplified for the reconstructed orbital and KS canonical orbital that we used in this study. \textbf{(b)} and \textbf{(c)} summarize the results for $E_{QP}$ and $\Sigma_P^{sub}$ for hybridization suppressed orbitals on the NP (111)$_f$ sites. QP energy axis is the left and the axis for $\Sigma_P^{sub}$, the polarization contribution term from the substrate is the right. The stochastic errors in $\Sigma_P^{sub}$ are shown by the black lines that are smaller than the red circle size.} 
        \label{fig3}
    \end{figure}

For each FMO, we take the respective magnitudes of its projection onto the molecular and surface subspaces to account for their contributions to the self-energy, as:
\begin{equation}
\Sigma = |p|^2 \Sigma_{\rm mol} + \left(1 - |p|^2\right) \Sigma_{\rm sub}
\end{equation}
This simple relationship that assumes that the molecular and surface subspaces are clearly separated (i.e., disregarding the off-diagonal contributions) corresponds well to our numerical data for $\Sigma$ as a function of $|p|^2$. The fitted parameters $\Sigma_{\rm mol}$ and $\Sigma_{\rm sub}$ are in the SI. We see, in Fig.\ref{fig3} that the injection propensity varies linearly with the degree of hybridization, with a slope of $\Sigma_{\rm sub} - \Sigma_{\rm mol}$. Note that the dependence changes when we consider projection on HOMO ($|p_{\rm HOMO}|^2$) or the entire molecular subspace ($|p|^2$) as illustrated in Fig.\ref{fig3}; yet, the main conclusions are unaffected.  As ($|p| \to 0$) the FMO levels are shifted closer to the substrate Fermi level.
The results above clearly indicate that hybridization between the molecule and the surface substantially impacts the charge injection energies. Still, the hybridization alone does not clarify the role of surface plasmons. We will thus further disentangle the self-energy and inspect the contribution stemming \textit{purely} from the charge density fluctuations in the substrate. We employ the self-energy decomposition scheme\cite{romanova2020decomposition} and compute $\Sigma$ of the molecular and substrate subspaces using the localized molecular orbitals basis (obtained by the orbital reconstruction \cite{weng2021efficient}
). In this localized molecular basis, the hybridization of molecular states has been artificially ``suppressed''.

The results support the picture of a large QP energy variation; there is more than $1$ eV difference between the facet center (111)$_{f1}$ and the facet edge (111)$_{f3}$ cases. (Fig.\ref{fig3}) As in the hybridized basis result, the HOMO QP energy is closest to the Fermi level when the molecule sits at the (111) facet center. In contrast, the hybridized and ``hybridization-suppressed'' results differ for the LUMO, whose energy shifts towards the Fermi level as the molecule approaches the NP edge. In the absence of molecule-substrate orbital hybridization (the latter case), the propensity for direct injection thus trends together with the typical strength of local electric field enhancement when approaching to the nanostructure edge that has been studied before.\cite{christopher2012singular,seemala2019plasmon} This observation is further illustrated in Fig.\ref{fig3}c, showing that the electron injection barrier decreases by nearly $1.3$ eV from (111) facet center (111)$_{f1}$ to the edge (111)$_{f3}$.
This plasmonic effect is captured by the variation in the correlation self-energy term, which contributes $66\%$ to the QP energy difference between (111) facet center and the edge, further motivating our use of MBPT techniques. In practice, however, this trend is significantly altered by hybridization.

As a final note, we remark that the DFT orbitals have limited physical significance and suffer from over-delocalization \cite{mori2008localization,cohen2008insights,kim2013understanding,johnson2013extreme}, yet, they provide physical insight into the role of hybridization on the QP energies. A deeper analysis of this effect will require the calculation of Dyson orbitals, which requires diagonalization of the full QP Hamiltonian matrix\cite{Kaplan2015,caruso2013self,rusakov2016self}. Such a step requires new developments in order to be computationally affordable for systems of the sizes discussed in this work. 

In summary, we conducted a first-principles study of charge injection into molecules on a plasmonic gold surface using many-body perturbation theory. The molecular quasiparticle energies take into account the substrate-induced renormalization via coupling to charge density fluctuations (plasmons). The injection barrier varies substantially depending on the precise position on the nanoparticle surface; such variation is not seen in the mean-field (DFT) calculations and clearly illustrates the need for the inclusion of more elaborate (non-local and dynamical) correlation energies. The variation is on the order of hundreds of meV, and it significantly impacts the molecule's propensity to accept hot holes or electrons. We identify the degree of hybridization as one of the decisive effects on identifying the ``hot spots'' in terms of the vertical excitation besides the plasmon coupling factor. In fact, for the lowest electron injection energy, the hybridization overshadows the effect of the coupling purely to the surface charge fluctuations.   It implies that in the structural design seeking the efficiency enhancement of plasmonically driven reactions should tune the chemical properties of the substrate to drive a higher degree of hybridization. While the pure substrate-driven renormalization (in the absence of hybridization) is strongest at the nanoparticle edges, the molecular states more readily delocalize over the NP substrate near the facet centers and completely alter the energy landscape. The direct injection mechanism needs to overcome a relatively large energy barrier, requiring a strong non-equilibrium hole/particle distributions well below/above the Fermi levels. Even then, the energy variation for hot holes is significantly larger than for the injection of hot electrons. In practice, the charge injection may be spatially limited to only very narrow regions of the nanoparticle facets (e.g., near the center of 111 surfaces). This work underlines the need for the inclusion of nonlocal and dynamical correlation effects and motivates further research bringing the condensed matter theoretical methods into the study of chemically relevant molecular excitations on (optically active) surfaces. 

\begin{acknowledgement}
The authors thank Dr. Mariya Romanova and Dr. Guorong(Gwen) Weng for their technical support. This work was supported by the NSF CAREER award (DMR-1945098). Use was made of computational facilities purchased with funds from the National Science Foundation (CNS-1725797) and administered by the Center for Scientific Computing (CSC). The CSC is supported by the California NanoSystems Institute and the Materials Research Science and Engineering Center (MRSEC; NSF DMR 2308708) at UC Santa Barbara.\
\end{acknowledgement}

\begin{suppinfo}
\ce{CO_2} geometry optimization on Au; Au substrates preparation; Hybridized single-particle state calculations; Two-state fitting of self-energy on the Au$_{260}$ NP; Decomposition of self-energy calculations.
\end{suppinfo}
\bibliography{plasmonics}

\end{document}